\documentclass[a4paper,12pt]{article}
\usepackage{amsmath,amsthm,amssymb,bbm,color}
\usepackage{braket}
\usepackage{dsfont}

\usepackage[english]{babel}
\usepackage{graphicx}

\title{\bf Noisy effects in interferometric quantum gravity tests}

\author{F. Benatti$^{a,b}$, 
R. Floreanini$^{b}$,
S. Olivares$^{c,d}$
E. Sindici$^{e}$,
\\
\\
\small ${}^a$Dipartimento di Fisica, Universit\`a di Trieste, 
34151 Trieste, Italy\\
\small ${}^b$Istituto Nazionale di Fisica Nucleare, Sezione di Trieste,
34151 Trieste, Italy\\
\small \hskip -0.8cm ${}^c$Quantum Technology Lab, Dipartimento di Fisica, Universit\`a degli Studi di Milano, 20133 Milano, Italy\\
\small ${}^d$Istituto Nazionale di Fisica Nucleare, Sezione di Milano, 20133 Milano, Italy\\
\small ${}^e$SUPA and Department of Physics, University of Strathclyde, Glasgow G4 0NG, UK
}

\date{\null}

\begin{document}

\maketitle

\begin{abstract}
\noindent
Quantum-enhanced metrology is boosting interferometer sensitivities to extraordinary levels, up to the point where table-top experiments have been proposed to measure Planck-scale effects predicted by quantum gravity theories. In setups involving multiple photon interferometers, as those for measuring the so-called holographic fluctuations, entanglement provides substantial improvements in sensitivity. Entanglement is however a fragile resource and may be endangered by decoherence phenomena. We analyze how noisy effects arising either from the weak coupling to an external
environment or from the modification of the canonical commutation relations in photon propagation may affect
this entanglement enhanced gain in sensitivity.
\end{abstract}

\vskip 1.5cm

\section{Introduction}

Most approaches to quantum gravity, either effective or fundamental, generally predict the appearance of
non-standard phenomena at the Planck scale, due to the ``foamy'' structure of spacetime.%
\footnote{The original idea that at Planck scale quantum fluctuations of the space geometry 
could destroy the smoothness of the space-time manifold has been
introduced in \cite{Wheeler} and, since then, further discussed by many authors 
({\it e.g.} see \cite{Hawking}-\cite{Hagar});
for recent reviews and further details, see \cite{Amelino, Plato} and references therein.}
It is however difficult
in general to estimate how these necessarily tiny disturbances could become visible in an
actual experimental setup. 

Interferometric apparata have emerged
as the most suitable setups for such kind of analysis, in particular those made 
of two identical photon interferometers: it has been shown that
using a couple of correlated interferometers in specific configurations may allow
to efficiently distinguish quantum gravity effects from other spurious signals \cite{Hogan1,Hogan2,Hogan3}.
%\textcolor{red}{In this scenario, these effects arise from the non-commutativity at the Planck scale
%($l_p \simeq 1.6\times10^{-35}$~m) of the position and momentum variables in different directions:
%this is the so-called ``holographic noise'' \cite{Hogan1}.}
The extreme sensitivities that these apparata need to reach in order
to actually measure these minuscule effects is nevertheless challenging
\cite{Holometer,Chou1,Chou2}.
\par
As in gravitational wave detectors \cite{Abbott,Grote,Chua}, quantum metrological methods may
be employed to enhance the sensitivity of such quantum gravity detectors \cite{Giovannetti}.
On the one hand, the use of nonclassical states, such as single- and two-mode squeezed states
can be a resource to enhance the sensitivity of optical interferometers \cite{Abadie}, allowing, 
at least in principle, to reach the so-called Heisenberg limit \cite{Sparaciari1,Sparaciari2}. 
On the other hand,
it has been shown that in the particular setup using two
photon interferometers, by feeding them with quantum correlated (entangled) initial
photons the overall sensitivity of the device may be dramatically enlarged,
at least in an ideal situation \cite{Genovese,Genovese2}.

Quantum entanglement is however a fragile resource, that can be endangered by various decohering
phenomena: therefore, it is of utmost importance to investigate to what extent the entanglement
enhanced sensitivity of the apparatus is robust against external noise.
Indeed, an interferometer is never completely isolated form the external environment,
which is in general a source of decohering phenomena. Furthermore, 
many fundamental theories predict various kinds of spacetime noncommutativity
at the undermost, basic level \cite{Connes}-\cite{Suijlekom};
these phenomena can affect the propagation of the photons
inside the interferometers through a modification of the canonical commutation relation,
leading to further noisy phenomena. All these unwanted effects
may reduce the enhancement in sensitivity obtained by feeding the apparatus
with highly non-classical, entangled light.

The general theory of open quantum systems \cite{Alicki}-\cite{Chruscinski}, 
{\it i.e.} systems in weak interactions with external baths,
can be used to estimate the effects produced by the external environment in the double interferometer
apparatus. In this framework, the propagation of the photons inside the experimental setup
is described by a quantum dynamical semigroup, generalizing the familiar unitary dynamics.
On the other hand, the existence of a minimum length, as predicted by most
theories based on noncommutative geometry \cite{Seiberg,Madore,Suijlekom}, 
may lead to a generalized uncertainty principle and as a consequence 
to a modification of the bosonic canonical commutation relations
the photon mode operators obey. 

In the following, we shall discuss in detail how the sensitivity enhancements 
provided by the use of entangled photons is affected by the presence of both sources of ``noise''.
In particular, we shall estimate how large the effects of these decohering phenomena should be
in order to spoil the enhancement in sensitivity when detecting quantum gravity effects
obtained through the use of quantum metrological methods.

\section{Detecting holographic fluctuations with entangled photons}

Photon interferometers are among the most accurate devices for detecting tiny
effects induced inside the apparatus by external perturbations. In the specific case of quantum gravity,
it has been predicted that the ``foamy'' structure of spacetime 
at the Planck scale may result in a noncommutativity of spatial coordinates, leading in turn to optical path-length differences between the two arms of the interferometer; the resulting
phase shifts have been named {\it holographic fluctuations} \cite{Hogan1}.

As mentioned above, this new kind of fluctuations cannot be detected by a single interferometer:
it is nearly impossible to isolate holographic fluctuations from other spurious signals,
even using extremely sensitive setups as gravitational antennas.
Therefore, specific configurations have been actually
designed and built in order to measure the accumulated phases coming from
holographic noise \cite{Holometer,Chou1,Chou2}.

\begin{figure}[t!]
\begin{center}
\includegraphics[scale=0.5]{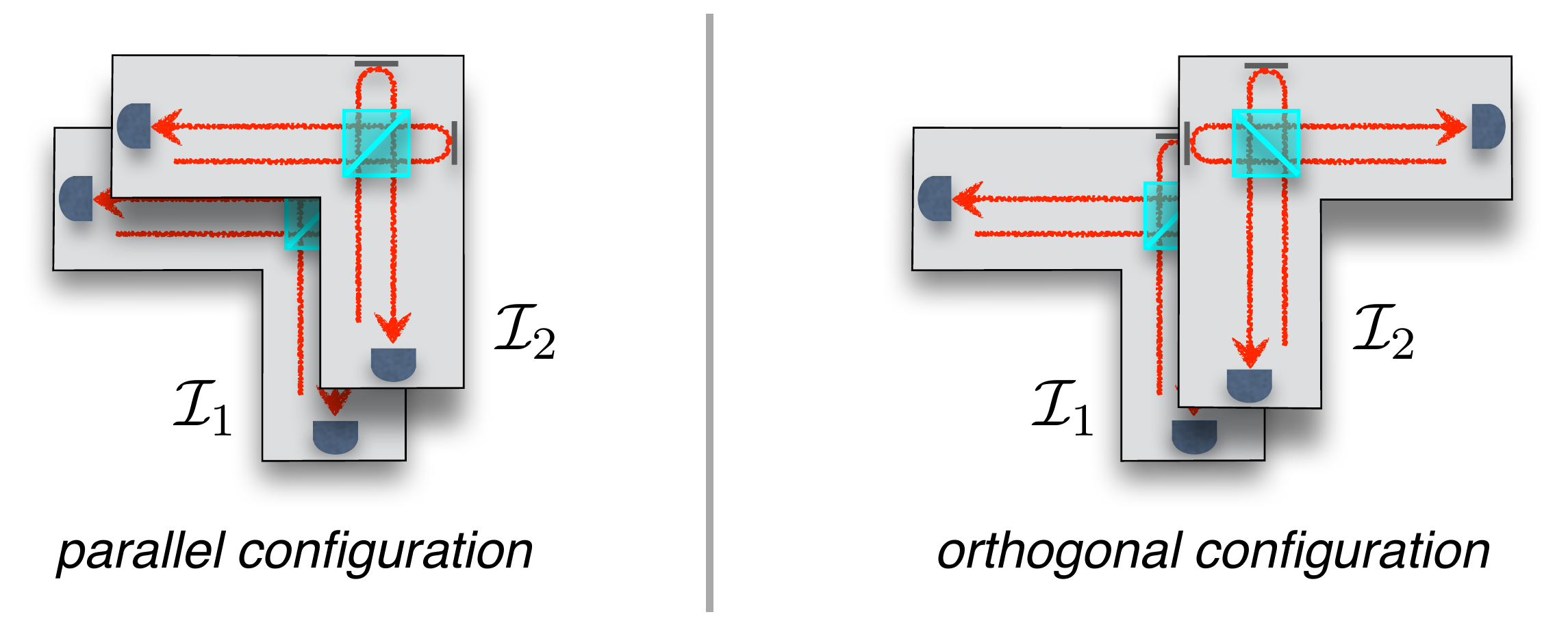}
\end{center}
\vspace{-0.7cm}
\caption{\small
Left: when two interferometers ${\cal I}_1$ and ${\cal I}_2$ are in the 
parallel configuration, they display correlated holographic fluctuations. Right:
in the orthogonal configuration, the correlation in holographic fluctuations vanishes.
More details about the interferometers are given in Figure~\ref{fig1}.}
\label{fig:P:O}
\end{figure}
If two slightly displaced parallel interferometers occupy overlapping spacetime volumes, then they display 
correlated holographic fluctuations (see the left panel of Figure~\ref{fig:P:O}, ``parallel configuration'');
on the other hand, by rotating one of the interferometers by $90^\circ$
degrees, so that one arm of the first interferometer becomes antiparallel to the one of the other, spacetime
overlapping is precluded and, as a consequence, the correlation in holographic fluctuations vanishes
(see the right panel of Figure ~\ref{fig:P:O}, ``orthogonal configuration'').
The second configuration can thus be taken as a reference measurement for the background signal, 
that, once subtracted from the outcome of the first configuration, should allow detecting the 
quantum gravity induced holographic fluctuations, provided sufficient sensitivity and statistics are achieved. 

The principal intrinsic limitation in achieving high accuracies in phase determination 
in such double interferometric devices is due to the shot noise limit. This limitation can be in part
circumvented by feeding the apparatus with nonclassical light. Indeed, as in the case
of more standard gravitational wave interferometry \cite{Abbott,Grote,Chua}, also in the case of 
holographic fluctuation measurements the use of squeezed light 
instead of classical coherent one would allow reaching higher sensitivities in phase estimation.

\begin{figure}[t!]
\begin{center}
\includegraphics[scale=0.4]{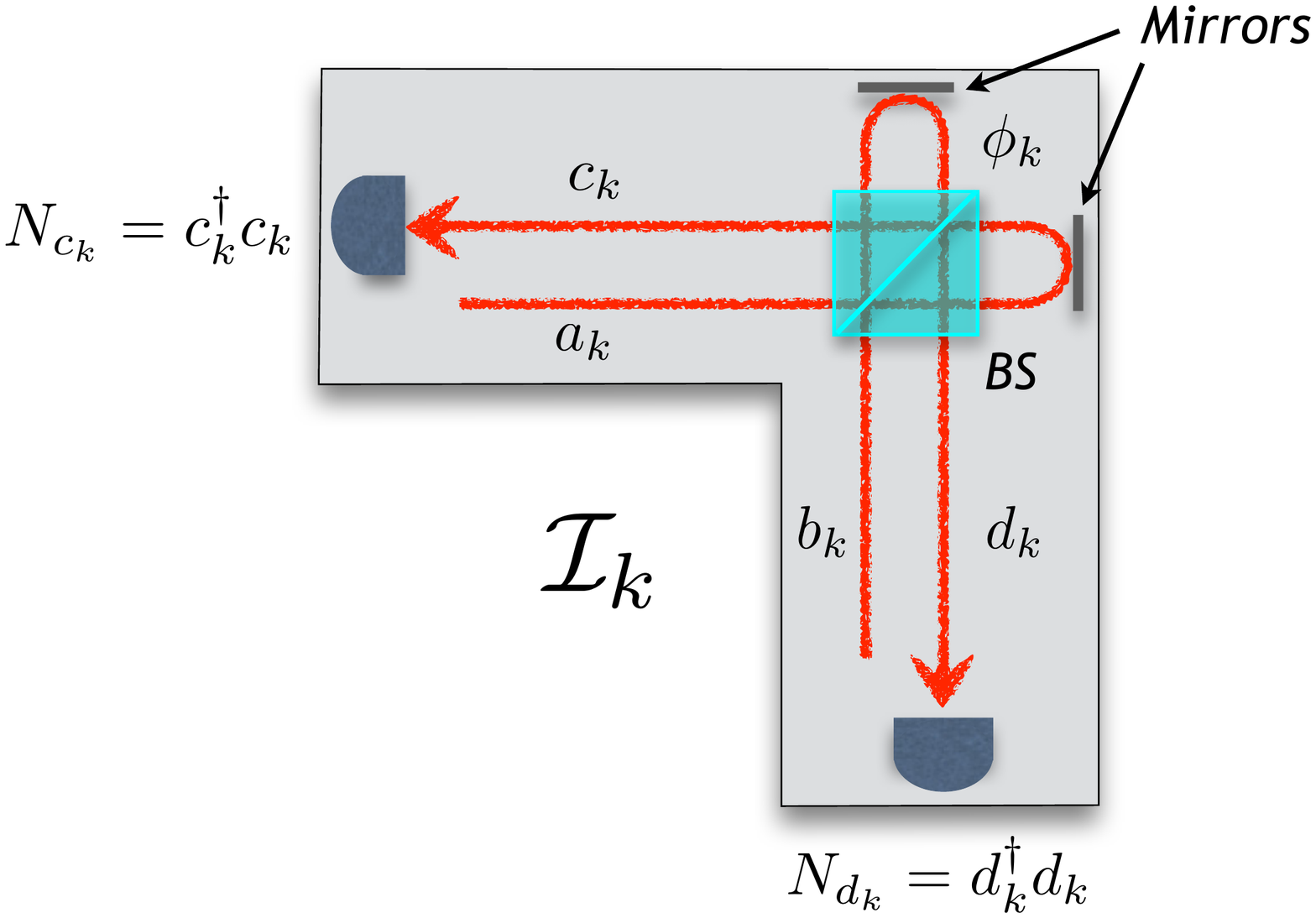}
\end{center}
\vspace{-0.7cm}
\caption{\small Schematic configuration of the $k$-th interferometer, $k=1,2$, 
with input $a$, $b$ and output $c$, $d$ port modes; BS represents the beam splitter,
while the detectors at the output ports measure the photon number $N_c$ and $N_d$.}
\label{fig1}
\end{figure}

However, the major breakthrough in sensitivity enhancement for the measurement of holographic fluctuations was shown to be brought in by feeding the double interferometer with suitable quantum correlated photons \cite{Genovese}.
The considered setup is made of two identical Michelson-like interferometers, labelled ${\cal I}_k$, $k=1,2$
(see Figure \ref{fig1}). The inputs fields of the interferometers
are described by the creation and annihilation mode operators
$a_k^\dagger$, $a_k$ and $b_k^\dagger$, $b_k$, $k=1,2$, obeying the standard bosonic commutation relations,
$[a_j^\dagger,\, a_k]=\delta_{jk}$, $[b_j^\dagger,\, b_k]=\delta_{jk}$; they are combined into a beamsplitter
giving rise to the output mode operators $c_k^\dagger$, $c_k$ and $d_k^\dagger$, $d_k$, respectively. The number of photons
in the output ports, $N_{c_k}=c_k^\dagger\, c_k$ and $N_{d_k}=d_k^\dagger\, d_k$, are measured
by means of two photodetectors. As previously mentioned, quantum gravity effects induce
an optical path length difference in the interferometers and therefore a phase shift $\phi_k$,
so that the relation between input and output modes is given by:
\begin{eqnarray}
\label{bs1}
&&c_k(\phi_k)=a_k \cos\left(\phi_k/2\right) + b_k \sin\left(\phi_k/2\right)\\
\label{bs2}
&&d_k(\phi_k)=b_k\cos\left(\phi_k/2\right) - a_k \sin\left(\phi_k/2\right)\ .
\end{eqnarray}
The configuration in which the two interferometers essentially overlap, having the corresponding arms aligned,
will be named \emph{parallel} ($\|$), while the second configuration in which one interferometer is rotated with
respect to the other, leading to two parallel and two antiparallel arms, will be called \emph{orthogonal} ($\perp$).

Let us now feed the $b$-ports of the two interferometers with photons in the same coherent state,
{\it i.e.}, $D_{b_1}(\mu)D_{b_2}(\mu) | 0 \rangle = |\mu\rangle |\mu\rangle$, where $D_{b_k}(\mu) = \exp(\mu\, b_k^\dag - \mu^*\, b_k)$
is the displacement operator of mode $b_k$, $k=1,2$, and $\mu\in \mathbb{C}$, 
while the $a$-ports with photons in an entangled squeezed state; in other terms,
the light entering the $a$-ports is quantum correlated between the two interferometers,
while the one entering the $b$-ports is not. The entangled state is obtained
by acting on the vacuum state with the two-mode squeezing operator 
\begin{equation}
S(\zeta)=\exp\left(\zeta a_1^\dagger a_2^\dagger - \zeta^* a_1 a_2\right)\ ,\qquad
\zeta=re^{i\theta}\ ,\quad r,\theta\in\mathbb{R}\ ,
\label{squeezing}
\end{equation}
giving rise to the two-mode squeezed vacuum state (the so-called twin-beam state):
\begin{equation}
\ket{\mathrm{TWB}}=\frac{1}{\cosh(r)}\sum\limits_{n=0}^\infty \big[\tanh(r)\big]^{n}\ e^{in\theta} \ket{n,\,n},
\label{twin-beam}
\end{equation}
where $|n,\, m\rangle$, $n,m\in\mathbb{N}$, are standard two-mode Fock states. This state
consists of a superposition of paired states with equal number of photons in each mode; as a result, 
it is a null eigenstate of any moment of the photon number difference operator, namely:
\begin{equation}\label{eq:property}
\big(a_1^\dagger a_1 -a_2^\dagger a_2\big)^p \ket{\mathrm{TWB}}=\,0\,, \quad \forall p \in {\mathbbm N} .
\end{equation}
In order to observe correlated phase-dependent fluctuations, one needs to study the behavior of
an observable which depends on both phases $\phi_k$, $k=1,2$; a convenient choice is given by 
the photon number difference
at the output $c_k$ ports of the interferometers, namely \cite{Genovese, Genovese2}: 
\begin{equation}
\Delta N(\phi_1,\phi_2)=\big[N_{c_1}(\phi_1)-N_{c_2}(\phi_2)\big]^2\ .
\label{4}
\end{equation} 
In addition, the holographic fluctuations are expected to be a stochastic process 
and therefore, in order to obtain averages to be compared with experimental outcomes,
the expectation of $\Delta N$ over the output photon states, hereafter indicated by $\langle \Delta N\rangle$,
needs to be further averaged over an appropriate probability distribution $ f_\alpha(\phi_1,\phi_2)$:
\begin{equation}\label{eq:expect}
\mathcal{E}_\alpha\left[\Delta N(\phi_1,\phi_2)\right]=\int \mathrm{d}\phi_1 \mathrm{d}\phi_2\,
\, f_\alpha(\phi_1,\phi_2)\, \braket{\Delta N(\phi_1,\phi_2)}\ ,\qquad \alpha=\|,\,\perp\ .
\end{equation}
One can show \cite {Genovese} that the holographic fluctuations are actually described by the following phase-shift correlations
\begin{equation}
\mathcal{E}_\|\left[\delta\phi_1\,\delta\phi_2\right]=\int \mathrm{d}\phi_1 \mathrm{d}\phi_2\, \delta\phi_1\, \delta\phi_2\, f_\|(\phi_1,\phi_2)\ ,
\end{equation}
where $\delta\phi_k=\phi_k-\phi_{k,0}$, $k=1,2$ are phase shift deviations from their corresponding
mean central value $\phi_{k,0}$. By making the reasonable assumptions that the distributions $ f_\alpha(\phi_1,\phi_2)$ 
have identical marginals and uncorrelated phase noise in the $\perp$ configuration,
one can relate this quantity to the differences of the two averages 
$\mathcal{E}_\|\left[\Delta N(\phi_1,\phi_2)\right]$ and $\mathcal{E}_\perp|\left[\Delta N(\phi_1,\phi_2)\right]$.
By expanding the explicit expressions of these expectations in series of $\delta\phi_k$ 
about the central values $\phi_{k,0}$, one finds, for small $\delta\phi_k$ \cite{Genovese}:
\begin{equation}
\label{7}
\mathcal{E}_{\|}\left[\delta \phi_1 \delta\phi_2\right]= \frac{\mathcal{E}_{\|}
\big[\Delta N( \phi_1,\phi_2)\big]- \mathcal{E}_{\bot}\big[\Delta N( \phi_1,\phi_2)\big]}
{\braket{\partial_{\phi_1}\partial_{\phi_2}\Delta N(\phi_1,\phi_2)}\vert_{\phi_k=\phi_{k,0}}}\ ,
\end{equation}
which holds when the denominator is nonvanishing.
We are interested in evaluating the uncertainty $\Delta\mathcal{E}$ with which this quantity can be
determined using the double interferometer apparatus, {\it i.e.}
\begin{equation}
\Delta\mathcal{E}=\Bigg[\frac{\mathrm{Var}_{\|}\left[\Delta N(\phi_{1},\phi_{2})\right]
+\mathrm{Var}_\perp\left[\Delta N(\phi_{1},\phi_{2})\right]}
{\braket{\partial_{\phi_1}\partial_{\phi_2}\Delta N(\phi_1,\phi_2)}^2\vert_{\phi_k=\phi_{k,0}}}\Bigg]^{1/2}\ ,
\end{equation}
where ${\rm Var}_\alpha\left[\Delta N(\phi_1,\phi_2)\right]$ are the variances of 
$\Delta N(\phi_1,\phi_2)$, in the two configurations $\alpha=\|,\,\perp$. Within the same approximation used above,
one finds that, to lowest order in $\delta\phi_k$:
\begin{equation}
\Delta\mathcal{E}\simeq\Bigg[\frac{2\,\mathrm{Var}_\|\left[\Delta N(\phi_{1},\phi_{2})\right]}
{\braket{\partial_{\phi_1}\partial_{\phi_2}\Delta N(\phi_1,\phi_2)}^2}\bigg\vert_{\phi_k=\phi_{k,0}}\Bigg]^{1/2}\ .
\label{9}
\end{equation}
This result was used in \cite{Genovese} to show that, by feeding the apparatus
with the two-mode squeezed vacuum state (\ref{twin-beam}),
one can obtain a substantial increase in sensitivity for holographic fluctuations detection with respect to an
analogous device using classical light. In particular,
in the special case $\phi_{1,0}=\phi_{2,0}=0$, 
the interferometers act like two completely transparent media,
as one can see from \eqref{bs1}, \eqref{bs2} by setting $\phi_k\equiv\phi_{k,0}=\,0$. 
Therefore, recalling the property (\ref{eq:property}),
the uncertainty $\Delta\mathcal{E}$ vanishes, while 
with coherent photon input states the shot noise limits the uncertainty to
$\Delta\mathcal{E}\geq \Delta\mathcal{E}_{\rm cl}\equiv\sqrt{2}/|\mu|^2$~\cite{Genovese}. 

This striking result holds only in an ideal setting, with interferometers working with perfect efficiency.
The robustness of these results against possible setup inefficiencies was also studied in \cite{Genovese,Genovese2}
by modelling them in terms of photon losses inside the apparatus.

In the following, we shall study how the evaluation
of $\Delta\mathcal{E}$ may be affected by the presence of a weakly coupled external environment
and by a gravity induced modification of the bosonic commutation relations obeyed by the photon modes.

\section{Noise induced by an external environment}

In a realistic scenario, the photons travelling inside the two interferometers
inevitably feel the presence of the surrounding, external environment, leading to noisy effects
that might endanger the accuracy in phase determination. 
In general, it is hard to estimate the form and magnitude of these unwanted effects due to
the complexity of the photon dynamics inside the apparatus; however,
in the specific situation at hand, the coupling between the photons and the environment can be assumed
to be very weak and in such a case their behaviour can be effectively described
using the well established theory of quantum open systems \cite{Alicki}-\cite{Chruscinski}.

Quite in general, the environment, which is made of an infinite number of microscopic degrees of freedom,
can be modelled as a free bosonic bath in equilibrium at a given inverse temperature $\beta$.
The total system, photons plus bath, can be initially prepared in a separable state of the form
$\rho\otimes\rho_\beta$, where $\rho$ is the density matrix describing the photon initial state,
while $\rho_\beta$ is the Gibbs density matrix describing the equilibrium state of the environment.
For rather generic (bilinear) interactions between photons and environment, 
in the limit of weak coupling, the reduced photon dynamics of the photons inside the interferometric setup,
obtained by tracing over the bath degrees of freedom,
can be described by a master equation in Kossakowski-Lindblad form \cite{Alicki}.

The relevant equation describing the time evolution
of the $a_1$ and $a_2$-mode photon states can then be cast
in the following form:
\begin{equation}\label{master-equation}
\frac{\partial \rho(t)}{\partial t} =-i[H,\,\rho]+ \sum\limits_{i,j=1}^{4} C_{ij} \left( V_j \rho V_i^\dagger
- \frac{1}{2} \left\lbrace V_i^\dagger V_j, \rho \right\rbrace \right),
\end{equation}
where $V_i$, $i=1,2,3,4$ represent the components of the four-vector $(a_1, a_1^\dagger, a_2, a_2^\dagger)$, while
$H=\omega_\gamma\sum_k a_k^\dagger\, a_k$ is the free photon Hamiltonian, with $\omega_\gamma$
the photon energy; in addition, $\{\ ,\, \}$ signifies anticommutation. The coefficient matrix $C_{ij}$, known as
Kossakowski matrix, contains the information about the environment. In the case of a free bath of bosons
at energy $\omega$ and temperature $T=1/\beta$, it can be taken of the following simple form \cite{Breuer,Gardiner}:
\begin{equation}
C_{ij} = \lambda
 \begin{pmatrix}
  1+M & 0 & 0 & 0 \\
  0 & M & 0 & 0 \\
  0  & 0  & 1+M & 0  \\
  0 & 0 & 0 & M
 \end{pmatrix}\ ,
 \end{equation}
where $M=\left(e^{\beta\omega}-1\right)^{-1}$ is the usual Boltzmann factor, while $\lambda$
is the photon-environment coupling constant.

In the master equation (\ref{master-equation}) 
one can distinguish two contributions: the first one is the standard Hamiltonian term leading to a
unitary evolution, while the other is responsible for noisy effects due to the presence
of the environment: the part containing the anticommutator produces dissipation, while the remaining
term leads to decohering effects. Due to these effects, the finite-time dynamics generated by 
(\ref{master-equation}) is no longer unitary, but of semigroup type, with composition
holding only forward in time.

As described in the previous Section, the states of the photons at the $a_k$ ports are prepared in an
entangled twin-beam state. In order to evaluate the environmental disturbances on phase estimation,
one has now to propagate in time this state according to the evolution equation (\ref{master-equation}) 
up to the time $\tau=4L/c$, where $L$ is the length of the interferometer arms 
and $c$ is the speed of light.

This can be more easily obtained by passing to a phase-space description, {\it i.e.} by introducing
the two-mode Wigner function corresponding to the photon state $\rho$:
\begin{equation}
W(z_1,z_2)=\frac{1}{\pi^2}\int \mathrm{d}^2\xi_1\, \mathrm{d}^2\xi_2\
e^{z_1\xi_1^*-z_1^*\xi_1 + z_2\xi_2^*-z_2^*\xi_2}\,
\mathrm{Tr}\left[\rho\, D_1(\xi_1)\, D_2(\xi_2)\right],\quad z_1,z_2\in \mathbb{C}\ ,
\end{equation}
where $D_k(\xi_k)=\exp(\xi_k a_k^\dagger- \xi_k^* a_k)$, $k=1,2$ are the so-called displacement operators. 
This description is completely equivalent to the one in terms of density matrices; in particular, it allows
computing expectation values of monomials in the mode operators $a_1$, $a_2$ by means of the following
general formula \cite{Glauber}:
\begin{align}
\nonumber
\mathrm{Tr}\big[\rho\,(a_1^\dagger)^{n_1} a_1^{m_1}(a_2^\dagger)^{n_2} a_2^{m_2}\big] &
=n_1!n_2!\left(-\frac{1}{2}\right)^{n_1+n_2}\iint \mathrm{d}^2 z_1 \mathrm{d}^2 z_2\\
&\times z_1^{m_1-n_1}z_2^{m_2-n_2}\, L_{n_1}^{(m_1-n_1)}(2|z_1|^2)\, L_{n_2}^{(m_2-n_2)}(2|z_2|^2)\, W(z_1,z_2),
\label{Glauber-formula}
\end{align}
where $L_n^{\alpha}(x)$ are the generalized Laguerre polynomials.

The master equation (\ref{master-equation}) generating the dissipative time evolution of the photon density matrix $\rho(t)$
becomes a Fokker-Planck equation for the corresponding Wigner function; explicitly, one finds:
\begin{equation}\label{Fokker-Planck}
\partial_t W_t(z_1,z_2)=\frac{\lambda}{2}\biggl[\sum_j\left(\partial_{x_j}x_j + \partial_{y_j}y_j\right)
+ (2M+1)\sum_j(\partial^2_{x_j}+ \partial^2_{y_j})\biggr] W_t(\alpha_1,\alpha_2)\ ,
\end{equation}
where $x_j$ and $y_j$ are the real and imaginary parts of $z_j$, $j=1,2$. 

The initial $a$-port photon state is the entangled twin-beam state (\ref{twin-beam}), and its corresponding Wigner function is of Gaussian
form. Since the equation (\ref{Fokker-Planck}) contains at most second order derivative, it preserves
its Gaussian form; indeed, the time evolved Wigner function takes the explicit form \cite{Olivares}:
\begin{equation}
W_t(z_1,z_2)=\frac{1}{4\pi^2\Sigma_+^2\Sigma_-^2}\exp\Bigg\lbrace-\frac{(x_1+x_2)^2}{4\Sigma_+^2}-\frac{(y_1+y_2)^2}{4\Sigma_-^2}-\frac{(x_1-x_2)^2}{4\Sigma_-^2}-\frac{(y_1-y_2)^2}{4\Sigma_+^2}\Bigg\rbrace\ ,
\end{equation}
where the functions
\begin{equation}
\Sigma_\pm=\frac{1}{2}\bigg(M+\frac{1}{2}\bigg)\big(1-e^{-\lambda t}\big)+\sigma_\pm e^{-\lambda t}\ ,
\end{equation}
give the explicit time dependence, while the coefficients $\sigma_{\pm} = e^{\pm {2 |r|}}$
contain the dependence on the initial squeezing parameter $r$ ({\it cf.} (\ref{squeezing})).

\begin{figure}[t!]
 \centering
   \includegraphics[width=15cm]{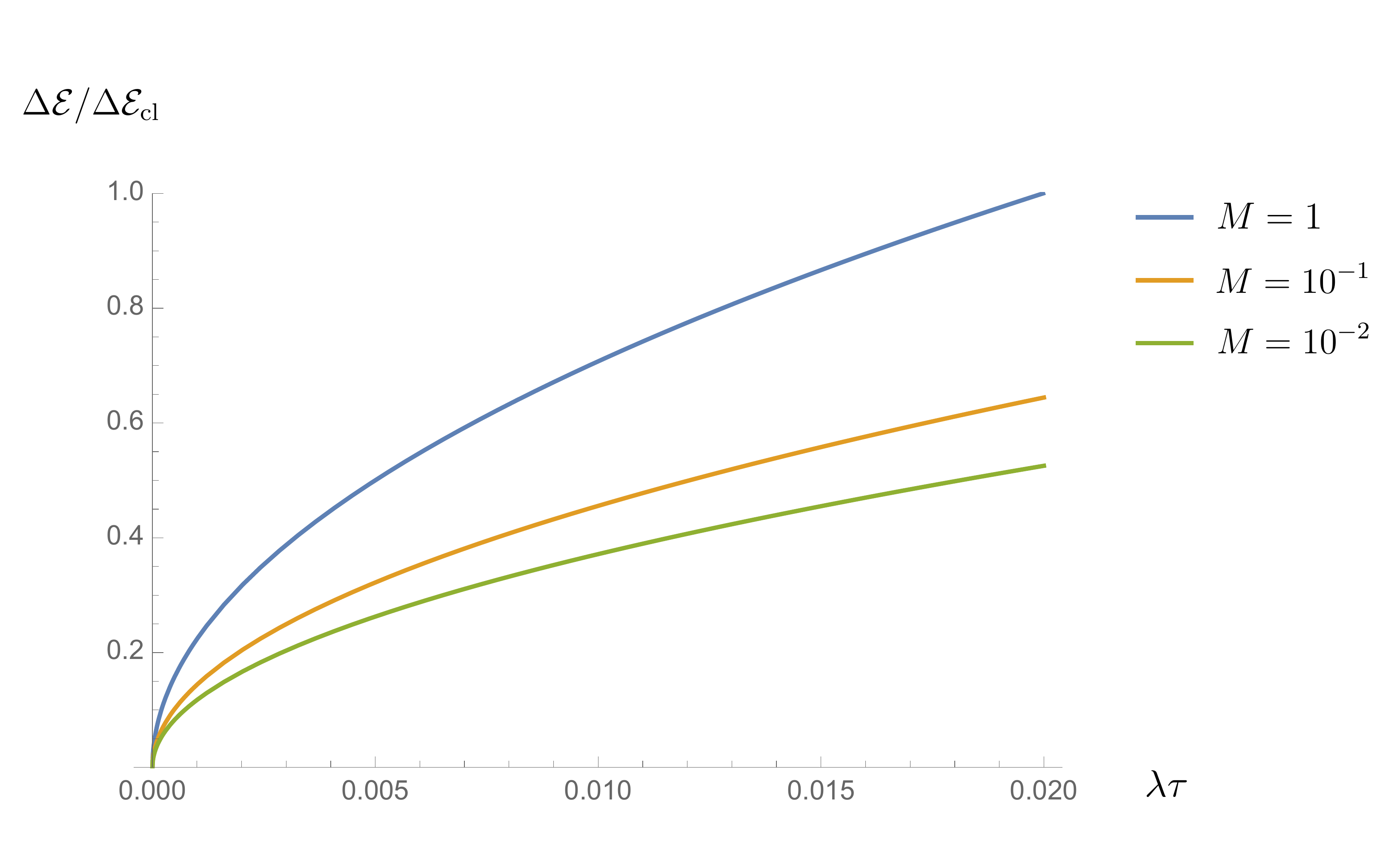} 
    \caption{\small Behaviour of the uncertainty, normalized to its classical value, 
in presence of an external bath, as a function of the coupling parameter $\lambda \tau$, for different 
values of the parameter $M$, with squeezing $r=2$.}
    \label{fig:diss}
\end{figure}

Using this result with $t=\tau$, the photon flight time inside the interferometers, 
and the general formula (\ref{Glauber-formula}), one can now evaluate how the uncertainty
$\Delta\mathcal{E}$ in the determination of the holographic fluctuations in (\ref{9}) 
is altered by the presence of the environment. As at the end of the previous Section, we shall consider
the case in which the central values of the phase shifts vanish, $\phi_{1,0}=\phi_{2,0}=0$,
so that any deviation from the ideal result $\Delta\mathcal{E}=\,0$ there obtained is
due to the noisy effects induced by the environment.
In this situation, recalling (\ref{4}) and (\ref{9}), the expression of the uncertainty
$\Delta\mathcal{E}$ reduces to:
\begin{equation}
\Delta\mathcal{E}/\Delta\mathcal{E}_{\mathrm{cl}}=
2\,\frac{\Big( \big\langle  \Delta \mathcal{N}^{\,4}\big\rangle 
-\big\langle \Delta \mathcal{N}^{\,2}\big\rangle^2 \Big)^{1/2}}
{\big\langle (a_1^\dagger + a_1)(a_2^\dagger + a_2) \big\rangle}\ ,\qquad
\Delta \mathcal{N}=a_1^\dagger a_1 - a_2^\dagger a_2\ ,
\label{18}
\end{equation}
where $\Delta\mathcal{E}_{\mathrm{cl}}$ is the uncertainty obtained feeding the apparatus with
classical coherent light.

The explicit evaluation of this ratio is cumbersome but straightforward,
and it amounts to the computation of integrals of the form (\ref{Glauber-formula}) 
with monomials up to order four \cite{Sindici}.
To lowest order in the small parameter $\lambda\tau$, one finds:
\begin{equation}
\Delta\mathcal{E}/\Delta\mathcal{E}_{\mathrm{cl}}\simeq \frac{8\, \sqrt{\lambda\tau}}{\sinh (2 r)} 
\Big[(2 M+1) \cosh (2r)-1\Big]^{1/2}\ .
\label{19}
\end{equation}
In general, also the coherent states $|\mu\rangle$ entering the other two ports of the
apparatus are affected by damping, so that they are modified by additional terms of order
$\lambda\tau$ or smaller.
However, their contribution
to the uncertainty involves in practice only the denominator of~(\ref{18}); 
since the numerator turns out to be proportional to $\sqrt{\lambda\tau}$,
one can compute the denominator in the zero-th order approximation,
{\it i.e.} with ordinary coherent states.

The behaviour of the ratio $\Delta\mathcal{E}/\Delta\mathcal{E}_{\mathrm{cl}}$ as a function of
the dimensionless parameter $\lambda\tau$ is reported in Figure 2, 
for different values of the parameter $M$.
As discussed before, this parameter describes the bath properties, and in particular it
contains the dependence on the bath temperature. However, imperfections in the preparation of
the initial twin-beam state could result in additional, effective ``thermal'' noise,
that can further contribute to $M$ \cite{dauria}. It is thus preferable to study the behaviour of the ratio 
in (\ref{19}) for different values of $M$ instead of directly the bath temperature, 
treating $M$ as an effective thermal parameter.
 
Although the presence of the bath makes now the uncertainty nonvanishing, 
the advantage of feeding the apparatus with entangled photon state is still apparent, provided the
couplings with the external environment is kept small. Notice that the uncertainty indeed
approaches zero in case of vanishingly small coupling $\lambda\tau$.
The behaviour of the uncertainty as a function of the squeezing parameter is instead
reported in Figure 3: one realizes that by increasing $r$ one can effectively
contrast the noisy action of the bath. 
According to these plots, the ratio $\Delta\mathcal{E}/\Delta\mathcal{E}_{\mathrm{cl}}$ appears to become
infinitely large for vanishing squeezing: this is due to the approximation used in deriving the formula in (\ref{7})
which ceases to be reliable for vanishingly small $r$, as its denominator becomes zero.%
\footnote{Only for a zero temperature bath ($M=0$) the expression in (\ref{19}) is still valid even for vanishing
squeezing: in this case, no advantage in sensitivity should be gained with respect to a ``classical'' apparatus,
as the the light entering all ports of the double interferometer is coherent, and indeed, one finds:
$\Delta\mathcal{E}\simeq \Delta\mathcal{E}_{\mathrm{cl}}$.}

The setup design proposed in \cite{Genovese} that uses entangled photons
appears therefore rather robust against environmental noise: 
the apparatus still retains a better sensitivity in holographic fluctuations determination
than the one attainable using classical coherent light.

\begin{figure}[t!]
 \centering
   \includegraphics[width=15cm]{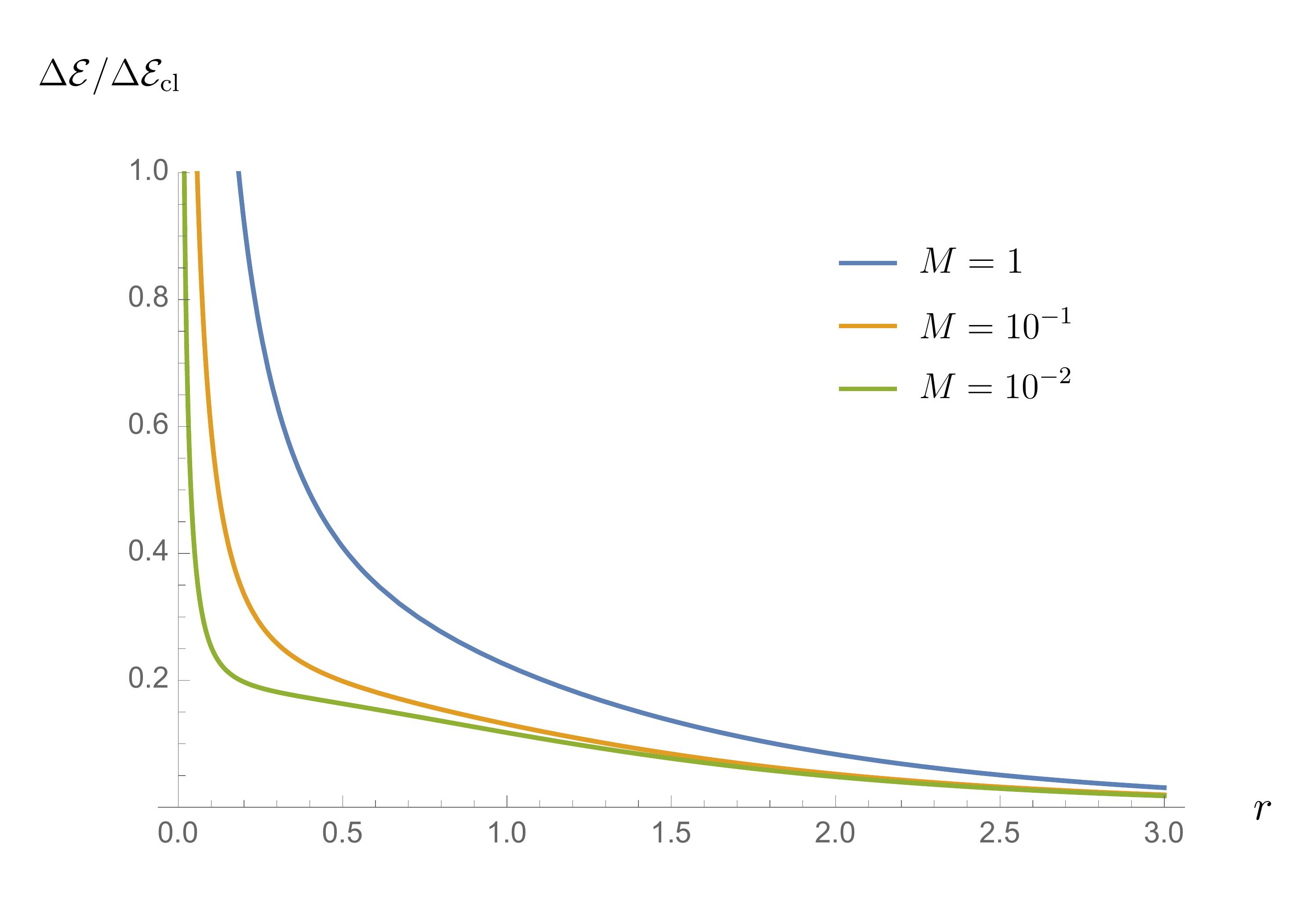} 
    \caption{\small Behaviour of the uncertainty, normalized to its classical value, 
in presence of an external bath as a function of the squeezing parameter $r$, for different 
values of the parameter~$M$, in a regime of weak coupling, $\lambda \tau=10^{-3}$.}
    \label{fig3}
\end{figure}

As a final remark, notice that the ``foamy'' structure of spacetime at the Planck scale
can itself effectively act as a noisy environment for the propagating photons \cite{Benatti2, Benatti3}.
In this case, on rough dimensional grounds, one can estimate the dimensionless coupling 
parameter $\lambda\tau$ to be suppressed by at least an inverse power of the Planck mass $M_P$,
{\it i.e.} $\lambda\tau\simeq \omega_\gamma/M_P$, with $\omega_\gamma$ 
the mean photon energy (see \cite{Benatti2,Benatti3} for further discussions).
For typical photon energy used in experiments ($\omega_\gamma\simeq 1\ {\rm eV}$)
and squeezing parameter $r\simeq 1$, the normalized uncertainty
$\Delta\mathcal{E}/\Delta\mathcal{E}_{\mathrm{cl}}$ is found to be as small as $10^{-15}$.
Therefore, the decohering effects generated by quantum gravity induced environments
can be safely ignored. However, as discussed in the following Section, other Planck scale phenomena 
can still influence the behavior of the travelling photons inside the interferometers
and therefore affect the estimation of the uncertainty $\Delta\mathcal{E}$.

\section{Noise induced by modified commutation relations}

As mentioned in the introductory remarks, many approaches to fundamental physics predict
the existence of a minimum length,
leading in turn to a modification of the usual canonical commutation relations.
Taking an effective approach, the most general extension
of the coordinate-momentum commutation relations involves additional terms \cite{Connes}-\cite{Suijlekom}:
\begin{equation}
\label{eq:modcommutators}
[x_i,p_j]=i\delta_{ij} + g_{ij}\ ,\qquad [x_i, x_j]=\ell_{ij}\ ,\qquad [p_i, p_j]=h_{ij}\ .
\end{equation}
As a result of this modification, also the behaviour
of the photons inside the interferometers, especially those that are prepared in a highly nonclassical,
entangled state, may be altered as well.
Although in general the quantities $g_{ij}$, $\ell_{ij}$ and $h_{ij}$ may themselves be functions 
of the coordinates $x_i$ and momenta $p_i$, we shall hereafter consider a simplified model
where only the $x$-$p$ commutation relations are modified by a constant contribution,
{\it i.e.}
\begin{equation}
\label{eq:modcommutators:CONST}
[x_i,p_i]=i(1 + \varepsilon)\,,
\end{equation}
while  $[x_1, x_2] = [p_1, p_2] = 0$, and $[x_i, p_j]  = 0$ for $i \ne j$.
By passing from the phase space to mode operators, one easily sees that 
the standard canonical commutation relations can be altered as follows
\begin{equation}
\label{eq:modccr}
[a_1,a_2]=\varepsilon,\quad [a_1,a_2^\dagger]=\varepsilon,\quad[a_i,a_i^\dagger]=  1+\varepsilon\ ,
\end{equation}
by the introduction of a real, adimensional, 
phenomenological parameter $\varepsilon$, assumed to be small $\varepsilon\ll 1$.%
\footnote{Assuming that the noncommutative effects originate at a fundamental energy scale $M_F$,
one can express this parameter as $\varepsilon=\alpha\,(\omega_\gamma/M_F)^\delta$,
with $\alpha$ an adimensional constant and $\delta=1,2$;
the value $\delta=2$ is favored by string theory models and black hole physics,
while $\delta=1$ can be motivated by more abstract group and algebraic considerations \cite{Tawfik}.
Experimental efforts try to set bounds on the parameter $\alpha$ in both of these scenarios,
using both astrophysical systems and table-top experiments ({\it e.g.} see \cite{Tawfik}-\cite{Pikovski}),
assuming for simplicity $M_F$ of order of the Planck mass.
A safe upper bound on the possible value of 
$\varepsilon$ that can be deduced from these
studies is of the order $10^{-1} - 10^{-2}$.}

The above modified commutation relations can be expressed in terms of standard 
mode oscillators $A_i$, $A_i^\dagger$, $i=1,2$,
\begin{align}
\nonumber
&[A_i,A_i^\dagger]=1\ ,\quad [A_i,A_j]=0\ ,\quad [A_i,A_j^\dagger]=0 \quad i\neq j\ ,\\
&A_i\ket{0}=0 \quad i=1,2\ ,
\label{eq:ccr}
\end{align}
through the following relations:
\begin{align}
\nonumber
a_1=&A_1\sqrt{1+\varepsilon} +\frac{\varepsilon}{2\sqrt{1+\varepsilon}}(A_2-A_2^\dagger )\ ,\\
a_2=&A_2\sqrt{1+\varepsilon}+\frac{\varepsilon}{2\sqrt{1+\varepsilon}}(A_1 + A_1^\dagger)\ ,
\label{eq:nua}
\end{align}
which indeed reproduce (\ref{eq:modccr}). It should be stressed that $A_i$, $A_i^\dagger$ 
are just auxiliary operators, useful for performing actual computations as they obey
standard canonical commutation relations; instead, photon states must now be constructed and
described through the commutators in (\ref{eq:modccr}).

Notice that the algebra generated by (\ref{eq:modccr}) does not admit a Fock representation,
{\it i.e.} a representation based on a lowest weight state, as defined by the condition $a_i |0\rangle=\,0$.
In such cases, one defines the vacuum state through the auxiliary $A$-modes as in (\ref{eq:ccr}).
As a consequence, the two-mode squeezing operator $S(\zeta)$, constructed with the $a$-modes as
in (\ref{squeezing}), no longer generates the twin beam state (\ref{twin-beam}) when acting on the vacuum,
rather a modified one $\ket{\mathrm{TWB}'}=S(\zeta)\ |0\rangle$.
Since the parameter $\varepsilon$ is assumed to be very small, it will be sufficient to compute the new state
to first order in $\varepsilon$.

The new input state for the $a$-ports of the apparatus can then be obtained by first expanding
\begin{equation}
\zeta a_1^\dagger a_2^\dagger - \zeta^* a_1 a_2= \mathcal{A}+\varepsilon \mathcal{B}\ ,
\end{equation}
with
\begin{align}
\nonumber
&\mathcal{A}= \zeta\, A_1^\dagger A_2^\dagger -\zeta^*\,A_1 A_2 \ ,\\
&\mathcal{B}=\zeta\Bigg\{ \frac{1}{2}\bigg[\Big(A_1^\dagger + A_2^\dagger\Big)^2
+A_1^\dagger A_1 - A_2 A_2^\dagger\bigg]\Bigg\} - {\rm h.c.}\ ,
\nonumber
\end{align}
and then using
\begin{align}
e^{\mathcal{A}+\varepsilon \mathcal{B}} &= e^{\mathcal{A}}
\left\{ 1 + \int_0^1 du\, \frac{d}{du}
\Big[ e^{-u\mathcal{A}} e^{u(\mathcal{A}+\varepsilon \mathcal{B})}\Big]
\right\}\,, \\
&\approx 
e^{\mathcal{A}}
\left\{ 1 + \varepsilon \int_0^1 du \, e^{-u\mathcal{A}}\,  \mathcal{B}\, e^{u\mathcal{A}}
\right\}\ ,
\end{align}
to compute to first order in $\varepsilon$ the action of the squeezing operator $S(\zeta)$ on the vacuum. 
Assuming for simplicity a real squeezing parameter $\zeta\equiv r\in{\mathbbm R}$, 
and recalling that:
$$
e^{-u\mathcal{A}}\, A_{1,2}\, e^{u\mathcal{A}} = \cosh(ru)\, A_{1,2} + \sinh(ru)\, A_{2,1}^\dag\ ,
$$
one finally gets that the modified input state becomes:
\begin{equation}
\ket{\mathrm{TWB}'}= \ket{\mathrm{TWB}}+ \varepsilon r \, e^{r A}\bigg\{\frac{1}{2}
\Big(A_1^\dagger + A_2^\dagger\Big)^2 -1\bigg\}\ |0\rangle\ ,
\label{eq:twb'}
\end{equation}
where, now, $\ket{\mathrm{TWB}} = e^{r\left(A_1^\dag A_2^\dag - A_1 A_2\right)} | 0 \rangle$;
the state $\ket{\mathrm{TWB}'}$ is a combination of entangled states.
\begin{figure}[t!]
 \centering
   \includegraphics[width=14cm]{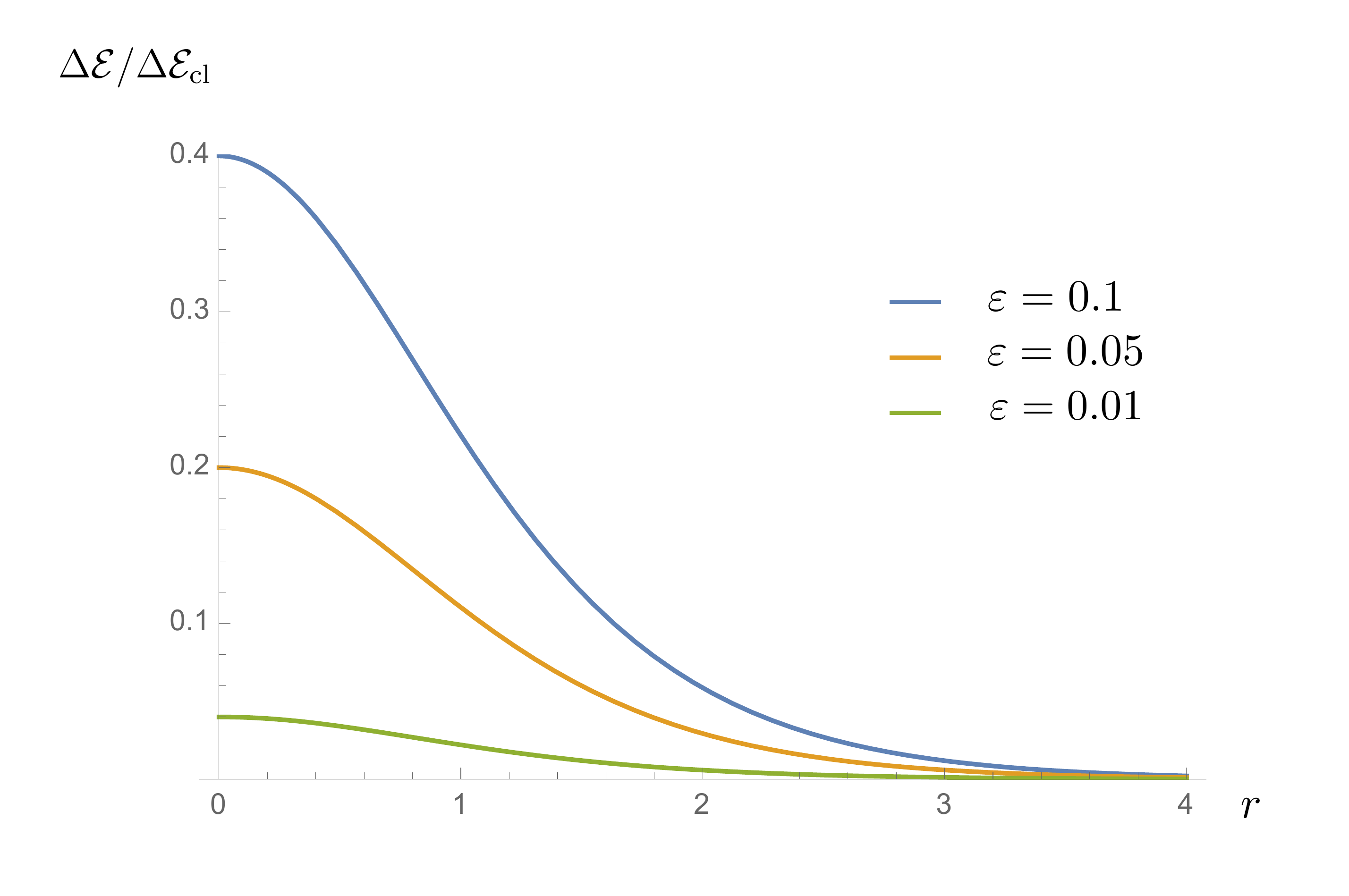} 
    \caption{\small Behaviour of the uncertainty, normalized to its classical value, 
in presence of modified photon-mode commutation relations, as a function of the squeezing parameter $r$, 
for various values of the deformation parameter $\varepsilon$.}
\label{fig:nocomm}
\end{figure}

\noindent
Using similar techniques and approximations, one can now evaluate
the uncertainty $\Delta\mathcal{E}$ in holographic fluctuations estimation modified by presence 
of the parameter $\varepsilon$. As in the case of environmental noise discussed in the previous Section,
we shall assume zero central values of the phase shifts, $\phi_{1,0}=\phi_{2,0}=0$,
so that the result (\ref{18}) still holds, since
as mentioned before, in this condition the interferometers work as completely transparent media. 
Indeed, (\ref{18}) is the result of algebraic manipulations and does not depend
on specific properties of the states.
As a result, also in this case,
a non vanishing $\Delta\mathcal{E}$ can only be ascribable to the Planck scale modified
commutation relations (\ref{eq:modccr}). 

The explicit calculation gives, to first order 
in~$\varepsilon$:
\begin{equation}
\label{eq:nocomvar}
\Delta\mathcal{E}/\Delta\mathcal{E}_{\mathrm{cl}}=\frac{8\, r\varepsilon}{\sinh(2r)}\ .
\end{equation}
Notice that also the coherent states entering the other two ports of the
apparatus should be defined using the modified mode operators in (\ref{eq:modccr}), so that,
to first order in $\varepsilon$, similarly to (\ref{eq:twb'}) one can write: $|\mu'\rangle=|\mu\rangle
+\varepsilon{\rm-correction}$. However, since they contribute only to the denominator of~(\ref{18}), 
and the numerator is proportional to~$\varepsilon$,
one can compute the denominator in the zero-th order approximation,
{\it i.e.} with ordinary coherent states.

The behaviour of this ratio as a function of the squeezing parameter $r$ 
is plotted in Figure~\ref{fig:nocomm}.
One can clearly see that the enhancement in sensitivity for the detection of holographic noise
due to the presence of entangled initial photons is still present, even for relatively large values
of $\varepsilon$.

\section{Concluding remarks and outlooks}

The spacetime non-commutativity at the Planck scale that most quantum gravity theory predicts
can in principle be detected using suitable photon interferometric apparata.
The idea is that the non-commutativity in space position can induce quantum fluctuations
on the optical components of an interferometer: 
these disturbances modify the length of the optical path of the photons traveling inside
the setup, causing a measurable change in the overall optical phase shift. This signal, dubbed holographic fluctuation,
is predicted to be extremely small, but might be in the reach of setups using two interferometers,
especially if fed with highly nonclassical, entangled light.

These conclusions hold for an ideal apparatus, perfectly isolated from its environment.
Instead, we have here analyzed to what extent the entanglement enhanced sensitivity in detecting
holographic fluctuations proves to be robust against decohering effects. In fact, the photons travelling inside the
interferometers inevitably interact with their environment, and this leads to noise
and dissipation; furthermore, Planck scale non-commutativity, whose effects we want to detect, 
may itself act as a decohering mechanism via a modification of the canonical commutation relations
obeyed by the photon mode creation and annihilation operators. 

We find that, if the coupling of the photons with the external environment is weak, a constraint in general very well
satisfied in common experimental conditions, and
the violations of the standard photon mode commutation relations are small, a phenomenologically sensible assumption,
the examined decohering effects will not be able to completely nullify the advantages brought 
in by the use of entangled light. 
In other terms, our results seem to confirm the validity 
of the approach employing quantum-enhanced metrology for detecting quantum gravity Planck scale effects.

%\eject

\section*{Acknowledgements}
The authors would like to thank M. Genovese, I. Ruo-Berchera and I. Degiovanni 
for useful and stimulating discussions.

\end{document}